\documentstyle[11pt]{article}
\begin{document}
 
 
\begin{center}
{\bf 'Galaxy Associations' as Possible Common Features\\[2mm] of
Galaxy Clusters}

\vspace{0.15in}

V.G.~Gurzadyan$^{1,2}$ and
 and A.~Mazure$^{1}$

\end{center}
\vspace{0.2in}
 
1. IGRAP, Laboratoire d'Astronomie Spatiale, Marseille, France;

2. Department of Theoretical Physics, Yerevan Physics
Institute,  Yerevan, Armenia (permanent address).

\vspace{0.2in}
 
{\bf Abstract.} The results of the analysis of the 
subclustering in a sample of ESO Nearby Abell
Cluster Survey (ENACS) galaxy clusters, with data 
complemented from other studies, 
are presented. The analysis is performed by using the 
S-tree method  
enabling one to study the hierarchical properties of 
clusters, namely by
the detection of the main physical cluster and of its 
subgroups.
The results indicate: (a) systematically lower genuine cluster velocity
dispersions than known from previous studies;
(b) existence of 2-3 subgroups in each cluster.
Due to certain properties of specific dynamical entities,
we denote these subgroups
as  {\it galaxy associations}, which can become an essential
challenge for formation mechanisms
of galaxy clusters.

\vspace{0.2in}

\section{Introduction
}

The study of the hierarchical subclustering in  clusters 
of galaxies 
is one among the important problems of observational 
cosmology. Since 
clusters of galaxies as distinct from stellar systems 
(star clusters, galaxies)
have characteristic time scales exceeding or comparable 
to the Hubble time,
their physical and dynamical properties should contain 
direct information
on the mechanisms of their formation and the early 
evolution of the
Universe.
 
Though the conclusions are not absolutely unambiguous 
with respect to
individual clusters, the
studies in general do support the existence of subgroups 
in clusters of
galaxies (e.g. West et al 1988; Bird 1994; West 1994; 
Escalera
et al 1994), supported also by X-ray data on
clusters (Moher et al 1993; Fabricant et al 1993; 
Bohringer et al 1994;
Grebenev et al 1995; Zabludoff and Zaritsky 1995; West 
et al 1995).
Substructuring of clusters is also anticipated by 
theoretical studies
(e.g. White 1976; Cavaliere, Colafrancesco \& Menci 
1992, Jing et al 1995). 

The ESO Nearby Abell Cluster Survey 
(ENACS) being a large homogeneous dataset of redshifts
of galaxies in clusters (Mazure et al, 1996,
Katgert at al 1996), provide an interesting
possibility to
investigate this problem. This dataset contains high 
reliability redshifts 
on 5636 galaxies in the direction of 107 clusters 
selected from the Abell-Corwin-Olowin (ACO) catalogue
with richness $R\geq1$ and mean redshifts $z\leq 0.1$.

The proper study of the profound properties of 
substructures, besides 
accurate observational data, requires also refined 
methods of statistical 
analysis.
Among others, well known methods are the correlation 
functions
- mainly two-point, but also occasionally using  higher 
order ones, 
the minimal spanning tree, topological measures, 
wavelets, etc (for review of these methods, see Gurzadyan and Kocharyan
1994a). 
These methods are using the positional information 
usually completed with redshift data, e.g. in the
case of wavelet technique 
(Slezak et al 1990; Escalera et al 1994; Grebenev et al 1995),
one can combine  2D wavelets 
for positions, 
and 1D ones - for redshift information

The present analysis is using the S-tree method
(Gurzadyan et al, 1991, 1994; Gurzadyan 
and Kocharyan 1994a) developed for the study
of the properties of the many dimensional nonlinear 
systems and is
essentially based on the concepts of the theory of 
dynamical 
systems.
In the context of galaxy clusters its
advantage is in the {\it self-consistent} use of both, 
the 
kinematical information - the
redshifts, and positional one - the 2-coordinates, as 
well as the 
data on individual observable properties of galaxies - 
the magnitudes.
The method has been already applied to the study of the 
substructures
of the Local Group, Virgo cluster, Coma cluster, Abell clusters
(Gurzadyan et al, 
1993; Petrosian et al 1997; Gurzadyan and
Kocharyan 1994a;
Gurzadyan and Mazure, 1996, 1997):
e.g. in the case of the Local Group, aside confirmation 
of the general 
picture known from other studies, some new associations 
between particular 
galaxies have been observed.

Here we report the results of the analysis of 10 Abell 
clusters - A119, A151,
A262, A539, A978, A1060, A2717, A3651, A3667, A3822.
The redshift data used are mainly those of ENACS,
complemented in some cases with other sources. Data 
for clusters A262, A539, A1060 are taken from 
literature.

The study of the substructuring properties of clusters 
can enable one to
investigate the
behavior of various important parameters, such as the 
velocity dispersion
variation with radius  (den Hartog and Katgert, 1996), 
or  the luminosity and 
morphological segregation in the subgroups. The latter 
problem was recently 
considered in the case of the core of Virgo cluster 
(Petrosian et al 1997),
analyzed again via the S-tree method and morphological 
segregation between the 
subgroups has been noticed. The importance of such an 
information is in its
direct consequences for cosmological theories.

Our aim here, however, is to concentrate only on two 
concrete points: the
estimates of genuine velocity dispersions of clusters 
and the typical
dynamical characteristics of substructures, since the 
advantage of the S-tree
method is precisely 
in the detection 
of small-scale structures. More comprehensive study of 
the results
of the dynamical substructuring for each individual 
cluster will be
performed elsewhere.

The results of our analysis show:

 {\it First}, that 1D velocity dispersions of
physical clusters are typically lower than those 
obtained
by other methods  (e.g. Zabludoff, Huchra and Geller 
1990; den Hartog and 
Katgert, 1996). If so, this fact should be considered as 
supporting 
cosmological models with $\Omega<1$.
This conclusion is close to that of (Bahcall and Oh 
1996), 
based on the use of Tully-Fisher distances of Sc 
galaxies;

{\it Second}, that clusters do contain
bound subgroups, with non-equal number of galaxies, but 
with typical
velocity dispersions in the range 100-200 km $s^{-1}$. 
The
most important is that  we have found some indication 
concerning their truncated Gaussian,
i.e. "box" like
velocity histograms. Simple dynamical considerations 
enable us to understand the
nature of this specific galaxy configurations, which, if confirmed, can have
essential
cosmological consequences. In view of their probable primordial nature,
is some analogy with the role of stellar associations in galaxies, we
call these dynamical entities {\it galaxy associations.}

We start with the description of the database in Section 
2, and a brief
summary of the S-tree technique in Section 3. The 
revealed substructures, i.e. the galaxy associations
in the sample of clusters of galaxies, are presented
in Section 4. The dynamical properties of galaxy associations are
analyzed in
Section 5. The results of the study and the main 
conclusions are
discussed in Section 6.

\section{The Data}

The following data were used in our analysis.
The choice of these 10 clusters was determined by the 
maximal completeness
in galaxies in the range of the  2 brightest magnitudes  
(for details see, Girardi et al 1997).

{\it A119}. ENACS, (Fabricant et al 1993).

{\it A151}. ENACS complemented with data of (Proust et 
al 1992).
The sample contains 142 galaxies, however for 8 of them 
(mostly
of high redshifts) no information
on magnitudes was available, therefore these objects 
were not included
into the analysis.

{\it A262}. (Sakai et al 1994; Gavazzi (unpublished)). 
Among 140 galaxies of 
the sample, no redshifts were available for 5 of them,
 and for 10 - no magnitudes 
(3 of them are in common); again these objects have been 
cancelled.

{\it A539}. (Girardi et al 1997). Among 113 galaxies of
the sample, 
 no magnitudes were available for 6 of them.

{\it A978}. ENACS. 

{\it A1060}. (Richter 1987; Lucey and Carter 1988; 
Zabludoff et al 1990).

{\it A2717}. ENACS, (Colless and Hewitt 1987;
Colless 1989).

{\it A3651}. ENACS. Contains data for 92 galaxies, but 
only for
80 of them magnitudes were available as well.  

{\it A3677}. ENACS. 

{\it A3822}. ENACS. 

\section{The S-Tree technique}

In this section we outline the key points of the S-tree 
technique (see Appendix),
referring for details to the original papers
(Gurzadyan et al, 1991,
1994, and to the monograph
(Gurzadyan and Kocharyan 1994a). One can also refer to
 the papers where these geometrical concepts have 
been firstly applied to
N-body gravitating systems (Gurzadyan and Savvidy,
1984, 1986).

The main idea of the S-tree method is based on the 
property
 of structural stability
(coarseness), well known in theory of dynamical systems,
enabling to reveal and study
the robust properties of nonlinear systems based on a 
limited amount of
information. As it was known before, gravitating systems 
are
exponentially unstable ones, and therefore, belong among 
systems 
possessing strong
statistical properties.

One application of this approach is the derivation of a 
formula
for the Hausdorff dimension depending on the dynamical 
properties
of the clusters of galaxies (Gurzadyan and 
Kocharyan 1991;
Monaco 1994). The S-tree method can  also be used 
for the
 reconstruction of
the 3D-velocities of clusters (Gurzadyan and Rauzy 
1996). 

The problem is formulated in a way to find out the 
correlation which should
exist between the parameters of an interacting N-body 
gravitating
system, i.e. between the velocity and coordinate 
components of the interacting
particles. Particles not satisfying the correlation 
found for a given
subgroup are considered as uninfluenced by (i.e. 
unbound to) that
subgroup. Thus,
the particles of set $A\subset N$ are considered to be 
influenced by each 
other if one can find out the ordinarity class   
\begin{equation}
\Phi_c : R^6 \times \Lambda \rightarrow R^3,
\end{equation}
and if there exists $c\in R^d$ such that for any $i\in 
A$
\begin{equation}
\Phi_c(x_i,v_i,\lambda_i)=0, 
\end{equation}
where $\lambda\in \Lambda$  describes the internal 
properties of the 
members of the system - the masses in our case.

The following two main concepts are basic ones for the 
S-tree method:
{\it the degree of boundness} of the various members of 
the 
system
(galaxies),
and the tree-diagram -- {\it S-tree} -- being 
constructed using the
boundness results and representing, therefore, the 
information on
the hierarchical substructure of the clusters.

The S-tree method offers
the possibility to use various affine parameters
while constructing the tree-diagram, such as the 
force of interaction, momentum,
potential, etc., as listed in (Gurzadyan and
Kocharyan, 1994).
However, the most complete information on the dynamics 
of the system
is extracted when the transition from the correlation to 
the degree of
boundness is realized by means of the simultaneous 
consideration of all 
particles
of the system, i.e. inquiring into the properties
of phase trajectories of the 
systems in $6N$-dimensional phase space with properly
defined measure.
This powerful method, well developed in the theory of 
dynamical
systems (see e.g. Arnold 1989),
enables one to reduce the problem of the dynamics of the 
N-body system to
the study of the geometrical properties of the phase 
space via the
estimation of the two-dimensional curvature:
\begin{equation}
K_{\mu \nu}=R_{\mu\nu\lambda\rho}u^{\lambda}u^{\rho},
\end{equation}
where $u$ is the velocity vector.

Then, the degree of boundness is obtained by means of 
the following
symmetrical matrix:
\begin{equation}
D_{ab}= \{ -K_{\mu \nu}, 0\};
\end{equation}
$$
\mu=(a,i), \nu=(b,j); 
a,b=1,2,..,N; i,j=1,2,3.
$$
Obviously, the tensor $K$ is containing 
self-consistently the information
both on the coordinates and velocities of all the 
particles of the system,
as well as on those individual properties of particles 
which are determining
their mutual interaction  -- the masses .

The results of the analysis of the hierarchical 
substructure
of the system, are represented via tree-diagrams -- {\it 
S-tree} -- as defined
in graph theory, by means of corresponding transition 
matrices $D_{ab}$:
\begin{equation}
\Gamma_{ab}=0 when D_{ab}< \rho,
\Gamma_{ab}=1 when D_{ab}> \rho.
\end{equation}
Obviously, the significance of the detection of the 
subgrouping hierarchy of
the system, as in any statistical problem, depends on 
the total number
of particles of the system $N$; numerical experiments 
described in the
above cited papers, show that the confidence level of 
the
substructures is $ > 90$ per cent if $N>30-35$; note, 
that 
all
substructures have the same confidence level, since they 
are
parts of the same interacting system.

As in previous studies of real clusters (Gurzadyan et al 
1993; Petrosian et al
1997)),
we presently used the assumption $ M = const L^n$ where 
$L$ is 
the luminosity, $M$ is the mass of galaxies, and
$n=1$ or $n=1/2$; the results are checked to be robust 
with respect to
these laws.

Let us briefly discuss the question of the 
correspondence of the S-tree method 
with
other statistical studies of clusters of galaxies; for a 
review
of the mathematical background of these various methods 
see (Gurzadyan and
Kocharyan 1994a).
Most thoroughly, the S-tree results have been
compared (together with Eric Escalera) with those of 
wavelets by applying  
this technique 
to the same initial sample of Abell clusters
(Girardi et al 1997).
We saw that the
main systems defined by both methods for most of the 
compared clusters are in
fair agreement. The S-tree method, however, enables also 
to
 reveal with the same confidence
level, substructures with smaller number of galaxies,
for which wavelets have a  limited sensitivity. The 
reason is obvious: 
for wavelets,
 a cD galaxy and its satellite have the same 
statistical weight in
attracting
the companions, while the S-tree method 
is using the information on the magnitudes,
and hence, on the masses of galaxies in a 
self-consistent way, which is
especially crucial while revealing  small-scale 
subgroups. In other
words, though wavelets uses also the redshift and
positional information, for example via 2D-wavelets for 
coordinates and
1D-wavelets for redshifts, both data are complementary 
but not
self-consistent; while the S-tree method inquiring into 
the
dynamics of an Hamiltonian system, is considering
the coordinates, velocities, and magnitude as subsets of
a single correlated set.
Thus, one can easily understand the difference of the 
S-tree technique
not only with respect to wavelets,
but also to other existing methods.

Among  methodical aspects of the use of the S-tree
method,
 let us mention also the role of
the magnitude completeness of the data sample in the 
final results.
Previous applications of the S-tree technique both to
toy models and to
real clusters (see e.g. Gurzadyan et al 1991, 1993, 
1994; Petrosian et al 1997)
had shown that, the main
substructure of the revealed system, as a rule, is 
determined by the
information on the brightest 2-3 magnitudes (depending 
on the total number of
galaxies in cluster, etc) galaxies of the system. In 
other words,
additional information on more fainter objects does not 
radically change
the revealed picture of substructuring. Indeed, faint 
objects can be either
projected more distant galaxies, which in any way would 
have 
correlation with the
parameters of the system,
or galaxies of much lower masses situated within the 
system.
Since the dynamics of the system is governed mainly by 
massive objects,
the corresponding magnitude limited samples will be 
reflecting
the genuine substructure of the clusters. The choice of 
the
sample of Abell clusters studied in present paper was 
done in view of these
conditions.

\section{The Substructure of the Sample of Abell 
Clusters: Existence of Galaxy Associations}

The S-tree analysis of the sample of Abell clusters 
defines: a) the main
physical system, as a system of interacting galaxies 
(MS), b) 
the subgroups of the main system. This includes the 
individual 
membership of galaxies in the
main system and in each subgroup. After that, the 
statistical analysis has been
performed in order to reveal the main physical 
parameters of
 each subgroup and of the
system.

The main results for the studied clusters are 
represented in Table 1.
It includes the total number of galaxies in the sample 
(T), in the revealed
main system (MS) and in the subgroups (1s, 2s, 3s), 
as well as 
their median velocity ($m$),
standard deviation ($\sigma$), skewness ($s$) and 
curtosis
($c$) of the velocity distribution. The redshift 
histograms are given in Figure 1.

\begin{table*}
\centering
\caption{Main parameters of the clusters: T denotes the 
total number
of galaxies in the sample, MS is the
main system, 1s, 2s, 3s are the corresponding subgroups, m is the median,
$\sigma$ is the standard deviation, s is the skewness and c is the curtosis
of velocity distribution}
\medskip
\begin{tabular}{lllllll}
\hline
\hline
A119    & T         & MS      & 1s       &  2s     & 3s  
   &  \\
\hline
N       & 142       & 109     & 53       &  18      & 13 
     & \\
m       &           & 13256   & 13152    & 13702    & 
12487   & \\
$\sigma$&           & 604     & 212      & 100      & 80 
     & \\
s       &           & 0.2     & -0.2     & 0.4      & 
3.1     & \\
c       &           & 0.3     &  -1.0    & -1.0     & 
-1.1    & \\
\hline
\end{tabular}
\end{table*}
\begin{table*}
\centering
\medskip
\begin{tabular}{llllll}
\hline
\hline
A151    & T         & MS      & 1s       &  2s     &  \\
\hline
N       & 134       & 59      & 23       & 16       &  
\\
m       &           & 15996   & 15845    & 16340    &  
\\
$\sigma$&           & 615     & 381      & 134      &  
\\
s       &           & -0.1    &  1.9     & 0.18     &  
\\
c       &           & -0.8    &  5.1     & -1.34    &  
\\
\hline
\end{tabular}
\end{table*}
\begin{table*}
\centering
\medskip
\begin{tabular}{llllll}
\hline
\hline
A262    & T         & MS     & 1s        &  2s      &  
\\
\hline
N       & 128       & 95      & 27       & 25       &  
\\
m       &           & 4864    & 4920     & 4600     &  
\\
$\sigma$&           & 481     & 120      & 116      &  
\\
s       &           & 2.84    &  0.41    & -0.5     &  
\\
c       &           & -0.93   &  -1.20   & -1.24    &  
\\
\hline
\end{tabular}
\end{table*}
\begin{table*}
\centering
\medskip
\begin{tabular}{llllll}
\hline
\hline
A539    & T         & MS      & 1s       &  2s      &  
\\
\hline
N       & 107       & 82      & 31       & 15       &  
\\
m       &           & 8737    & 8797     & 8279     &  
\\
$\sigma$&           & 627     & 120      & 143      &  
\\
s       &           & 0.52    &  0.23    & -0.1     &  
\\
c       &           & -0.35   &  -0.90   & -0.94    &  
\\
\hline
\end{tabular}
\end{table*}
\begin{table*}
\centering
\medskip
\begin{tabular}{lllllll}
\hline
\hline
A978    & T         & MS      & 1s       &  2s      & 3s 
     &  \\
\hline
N       & 73        & 58      & 12       &  20      & 6  
     & \\
m       &           & 16303   & 16771    & 16333    & 
16058   & \\
$\sigma$&           & 567     & 196      & 84       & 85 
     & \\
s       &           & 0.23    & 0.3      & -0.05    & 
-0.2    & \\
c       &           & -0.2    &  -1.6    & -1.4     & 
-1.9    & \\
\hline
\end{tabular}
\end{table*}
\begin{table*}
\centering
\medskip
\begin{tabular}{lllllll}
\hline
\hline
A1060   & T         & MS     & 1s        &  2s      & 3s 
     &  \\
\hline
N       & 105       & 82      & 20       &  44      & 13 
     & \\
m       &           & 3609    & 2957     & 3612     & 
4409    & \\
$\sigma$&           & 515     & 188      & 243      & 
112     & \\
s       &           & -3.82   & 0.1      & 0.2      & 
0.24    & \\
c       &           & -1.0    &  -1.57   & -1.33    & 
-1.4    & \\
\hline
\end{tabular}
\end{table*}
\begin{table*}
\centering
\medskip
\begin{tabular}{llllll}
\hline
\hline
A2717   & T         & MS      & 1s       &  2s      &  
\\
\hline
N       & 81        & 43      & 28       & 13       &  
\\
m       &           & 14814   & 14687    & 15125    &  
\\
$\sigma$&           & 338     & 184      & 126      &  
\\
s       &           & 0.32    &  -0.25   & -0.01    &  
\\
c       &           & -1.0    &  -1.1    & -1.5     &  
\\
\hline
\end{tabular}
\end{table*}
\begin{table*}
\centering
\medskip
\begin{tabular}{llllll}
\hline
\hline
A3651   & T         & MS      & 1s       &  2s      &  
\\
\hline
N       & 80        & 61      & 28       & 9        &  
\\
m       &           & 17943   & 17962    & 17544    &  
\\
$\sigma$&           & 531     & 157      & 91       &  
\\
s       &           & -0.05   &  0.3     & -0.4     &  
\\
c       &           & -0.6    & -0.7     & -1.44    &  
\\
\hline
\end{tabular}
\end{table*}
\begin{table*}
\centering
\medskip
\begin{tabular}{lllllll}
\hline
\hline
A3667   & T         & MS      & 1s       &  2s      & 3s 
     &  \\
\hline
N       & 113       & 90      & 23       &  10      & 24 
     & \\
m       &           & 16589   & 15847    & 17350    & 
16792   & \\
$\sigma$&           & 585     & 211      & 137      & 
208     & \\
s       &           & -1.4    & -0.3     & -0.1     & 
0.6     & \\
c       &           & -1.1    &  -1.3    & -1.5     & 
-0.85    & \\
\hline
\end{tabular}
\end{table*}
\begin{table*}
\centering
\medskip
\begin{tabular}{llllll}
\hline
\hline
A3822   & T         & MS      & 1s       &  2s      &  
\\
\hline
N       & 101       & 55      & 18       & 10       &  
\\
m       &           & 22883   & 22372    & 22958    &  
\\
$\sigma$&           & 537     & 273      & 64       &  
\\
s       &           & -0.05   &  1.0     & 0.2      &  
\\
c       &           & -1.2    &  0.4     & -1.5     &  
\\
\hline
\end{tabular}
\end{table*}
In addition to the parameters given in the Table 1, the 
analysis reveals
more features for each cluster. We briefly mention some 
of them.
For example, for A119 a fraction of the second system 
(12 galaxies) has been
indicated with a
mean redshift around 14,200 km $s^{-1}$, which probably 
has physical 
relation to the MS, so far as a weak correlation between 
them is observed;
for details see the paper by Gurzadyan and Mazure (1996) 
devoted entirely to
A119, containing also the comparison of S-tree results 
with the X-ray data, etc.
The sample of A151 also contains 36 galaxies of the
second system with a mean
redshift around 30,000 km $s^{-1}$, and fragments of 
other projected clusters.
In A2717, in the 1st subgroup, there is an indication of 
a core of 10 galaxies,
with median velocity 14527 km $s^{-1}$.
The initial sample of 81 galaxies, probably, contains 
fractions of a second 
system (11 galaxies).
A1060 also shows a fragment of a related second system 
(6 galaxies)
with mean redshift around 4800 km $s^{-1}$. 

These and other details about the clusters
revealed by the S-tree analysis  will be discussed
elsewhere.

What are the basic features of the substructuring 
paradigm
indicated by our analysis?

First, the existence of 2-3 subsystems is typical, at 
least for the
studied clusters.

Second, the magnitude of the velocity dispersions of 
the subsystems appears
to be around 100-200 km $s^{-1}$. As follows from  Table 
1,
the 4th momentum (curtosis) of the redshift distribution
shows a clear preference of negative sign.
This fact can be interpreted as a
tendency to have truncated Gaussian distributions.
For comparison, the 3rd momentum,
as expected, shows no preference in sign.

Although the conclusion on the existence of subgroups 
with truncated Gaussians -- {\it galaxy associations} -- is
strengthened by the fact that the S-tree method 
extracts
the subgroups and their parameters at the same 
confidence level as the whole
cluster, such a conclusion should be considered as a 
preliminary one.
However, if confirmed, this phenomenon can have deep 
physical
consequences, directly reflecting both -- 
the initial conditions of the cluster,
and its evolutionary paths, it makes sense to discuss at 
least
briefly some of its dynamical aspects.

\section{Why the galaxy associations can have ''box"
structures?}

Let us discuss  now, what can be the reason of truncated 
Gaussian, i.e.``box"
velocity distribution of the revealed substructures of 
the clusters.
Can it be an apparent effect, i.e. determined, say, by 
the data  incompleteness, especially for
  faint galaxies or  can this be a genuine 
feature of those configurations? 

We have a N-body gravitating system, which is a 
subsystem of
another gravitating system (main system), with larger N. 
What will be the 
velocity distributions of both systems? This problem is 
not new and, at 
first glance, is close to that of the dynamics of 
globular clusters 
within the galactic field. The situation in that case is 
basically
understood. So far as both systems (star clusters and 
host galaxies) are 
autonomous by their dynamics, and if
their corresponding relaxation time scales are less than 
the Hubble time,
then both
stellar velocity distributions should be close to 
Maxwellian; here we do
not discuss possible distortions of that law due to 
various effects, such
as evaporation of stars from the system, core collapse, 
dissolution and
formation of binaries, presence of massive black holes, 
etc. The observations
generally confirm this behaviour.

Quite different is the situation in the case of clusters 
of galaxies.
It is well known, that due to the difference in the 
order of N, many stellar
dynamical results cannot be directly applied to clusters 
of galaxies. In our
study of the substructures in clusters, we have one more 
demonstration
of this fact.

If the cluster of galaxies is a more or less isolated 
system, then 
depending on the total number of galaxies, its velocity
distribution can be close to Gaussian. This is what
is generally supported by observations.

However, the typical subsystems will not have Gaussian 
distribution, i.e.
distribution with ``wings", because of the following 
reason.
During the time scale
of the order of the dynamical time of the subsystem, 
$\tau_{dyn}$, the galaxies
situated in the wings of the velocity distribution 
should undergo
stronger attraction by the host cluster and have to 
escape
from the subsystem. The time scale of the  cut-off of 
the 
Gaussian distribution 
is $\tau_{dyn}$, while the time scale to recover the 
``wings" should be  $N^{\alpha}\tau_{dyn}$, where $N$ is 
the number of 
objects in the subsystem,
and  $1/3<\alpha<1$ is a positive        
number;
 its numerical value depends whether two-body or N-body 
interactions 
have the main role in that process (see Gurzadyan and 
Pfenniger 1994). 
Thus, the time scale of the cut-off will
always be shorter, than that of the recovering of the 
``wings", i.e. typically
one will observe only truncated Gaussian configurations 
-  {\it boxes}.
Note, that this dynamical process is somewhat different
from the tidal cut-off of globular clusters in the field 
of the host galaxy.

Another feature, which again is indicated by our 
analysis of observational
data, is the stability of standard deviations for the 
subgroups  velocities,
i.e. the stability  of the widths of the boxes.
The reason is again, understandable. The width of
boxes is essentially determined by the parameters of the 
host cluster, which
is cutting the ``wings".

Below, on a simple example we will illustrate this fact 
i.e. we will show in which way the properties of such 
subgroups 
can be determined by the parameters of the host cluster.

Since the number of galaxies within the subgroup is 
relatively small, 
instead of stellar dynamical technique one has to recall 
the methods of
celestial mechanics.

Assume that the galaxies in the subgroup are moving by 
elliptic orbits.
Then, the observed redshift interval $\Delta v= 
[v_{min},v_{vax}]$ should be 
determined by the maximal and minimal velocities of 
galaxies on the orbits:
\begin{equation}
v_{max}/v_{min}=(1+e)/(1-e),
\end{equation}
where $e$ is the eccentricity of the orbit.

Therefore, the cut-off of the redshift ``wings" will 
mean  the absence of
galaxies with
eccentricities higher than some $e_{max}$; i.e. galaxies 
with too elongated
orbits have to be captured by the host system and will 
not return to the system.
So, the problem comes to the determination of the 
population of systems with 
eccentricity cut-off at $e_{max}$.

In this formulation we have deliberately reduced the 
problem to the
classical one of
the binary stellar systems (Ambartzumian 1937), just  
treated by methods 
of celestial mechanics.

Indeed, consider the phase density of an $N$-body 
system, when the distribution 
function of galaxies is an arbitrary function only of 
energy 
\begin{equation}
dn=f(L_1,...L_N)\, dL_1,...dL_N\,dG_1,...dG_N\,
dH_1,...dH_N\, 
\end{equation}
$$
dl_1,...dl_N\,dg_1,...dg_N\,dh_1,...dh_N\,.
$$
where the Delaunay canonical elements $L_j$, $G_j$, 
$H_j$, $l_j$, $g_j$, and 
$h_j$ are defined as:
\begin{equation}
L_j= m_j( GM (\sum_{i=1}^{N-1}m_i/\sum_{i=1}^{N}m_i)\, 
a_j)^{1/2}\,, 
\qquad l_j=n_j(t-T_j)\,,                          
\end{equation}
$$
G_j= L_j(1-e_j^2)^{1/2} \,, \qquad 
g_j=\omega_j=\pi-\Omega_j\,,  
$$
$$      
H_j=G_j\cos i_j \,,\qquad h_j=\Omega_j\,,
$$                                                              
Here $a$ and $e$ are the major semi-axis and the
eccentricity of
the orbit of a galaxy, respectively,
$i$ is the inclination of its orbital plane, $n(t-T)$ is 
the mean
anomaly, $\omega$ is the angular distance of perihelion 
from the node,
$\Omega$ is the longitude of the ascending node, and $M$ 
is
the total mass of the system, $m$ is the galactic mass.

The advantage of considering the case of elliptic 
orbits is in the possibility to use the  Delaunay 
canonical formalism;
note, that $G$ and $H$ are positive, so far as $0<e<1$.

We are interested in the number of systems with cut-off 
of galactic orbits
at $e>e_{max}$:
\begin{equation}
n(G_j<G_{max})=8\pi^N \int_0^\infty f(L_1,...L_N) 
dL_1,...dL_N
\end{equation}
$$
\int_{0}^{L_1(1-e^2)^{1/2},...L_N (1-e^2)^{1/2}} 
dG_1,...dG_N\, 
G_1,...G_N \,,                
$$
since the integration over $G_j$ from 0 to 
$L_j(1-e^2)^{1/2}$ implies the 
variation of eccentricity from $e=1$ to $e_{max}$.

Therefore, the number of systems with cut-off of all 
orbits at $e_{max}$, is
\begin{equation}
n(e_j>e_{max})=A(1-e^2_{max}),
\end{equation}
where
\begin{equation}
A= const \int_{0}^{\infty}f(L_1,...L_N)L_1^2,...L_N^2\, 
dL_1,...dL_N\,.            
\end{equation}
i.e. the number of such systems crucially depends on 
$e_{max}$,
as, obviously, it was for the binary systems. 
From here one can conclude that the cut-off subsystems 
should be quite 
common within the host clusters, with the
cut-off parameter $e_{max}$, determined essentially by 
the parameters
of the host system. The marginal stability of
velocity dispersions around $\sigma \simeq 100-200$ km 
$s^{-1}$ is an indication
of this conclusion. Some proportionality of $\sigma$ on 
the number of galaxies
of subsystem can also be felt for some clusters (A119, 
A151, A262, etc),
which is again understandable.

One can even roughly estimate the range of
peculiar velocities of such configurations. 
Indeed, assume that the maximal
distance, ``perigee", of the elliptic orbit from the 
mass center, $r_{max}$,
does not exceed 
by the order of magnitude the mean distance between 
the galaxies
of the main cluster (otherwise the galaxy will be 
captured),
while $r_{min}$  should be of the order of the mean 
distance
between the galaxies in the subsystem.
Then, one can 
readily have: $e_{max}=1-2(R/R_0)(N_0/N)^{1/3}$, where 
$N_0$ is the number
of galaxies in the main cluster, $R_0$ and $R$
are the dimensions of the main system and of the 
subgroup, respectively.
If $e<<1$, one has
for their peculiar velocity interval: $\Delta v \simeq 
2v_0 e_{max} \simeq
(0.2-0.4) v_0$, 
($v_0$ is the mean velocity of main cluster) for the 
parameters of the above
studied clusters, i.e. in marginal agreement with the
obtained velocity range. Note, the weak proportionality 
of $\Delta v$
on the number of galaxies in the subsystem.

By this schematic analysis we aimed to demonstrate only 
the basic features of 
the galaxy associations, namely the typicalness of their
velocity cut-off and 
the crucial dependence on the parameters of the host 
cluster.
More refined theoretical studies are desirable to study 
their detailed
properties.

\section{Discussion.}

Thus, we have analyzed by using the S-tree method the 
substructure properties
of a sample
of clusters of galaxies, mainly issued form ENACS data, 
extracting with the same 
statistical significance level both the main cluster and 
its subgroups.

As a result, we have obtained the main parameters of the 
revealed clusters
and their subgroups. 
The 1D velocity dispersions obtained for the main 
systems are 
very close to those obtained by Bahcall and Oh (1996). 
This fact,
as discussed by these authors, is 
supporting the low $\Omega$ Friedmann Universe.
The hierarchical structure of the clusters revealed by 
the present analysis,
with different velocity dispersions, and hence, defining 
potential
wells of various scales,  
is in agreement with the hierarchical structure of the 
dark matter,
indicated by X-ray data (Ikebe et  al 1996).   

Among the basic properties of the subgroups  appears 
to be their non-Gaussian, 
namely, box-like 1D velocity distribution. Though,  any 
system with little 
number of gravitating bodies, will 
not in principle 
have ``smooth" Gaussian distribution, the box-like 
truncation,
if confirmed, can be a result of definite physical 
evolution. Indeed, simple
theoretical considerations enable to conclude that
box-like shape cannot  be due to statistical 
incompleteness or other
systematic errors, but most probably indicates the 
genuine property of 
these configurations -- {\it galaxy associations} -- 
namely, as subsystems,
whose part of galaxies
have been captured by the main cluster.    
Depending on the initial conditions and the paths of 
their evolution, i.e.
the degree of stripping during the evolution, the core 
of these
configurations can be close or far from the Gaussian 
cusp, though 
without ``wings". 

Also note little overlapping between the 1D velocity distributions
of the subgroups (Figure 1). Namely, this yields in average about 10
percent of members of other groups and 20 percent of members of the host
galaxy within the redshift interval of a given subgroup which are not
considered as its members.

This can be understood as follows: if the mutual bulk velocities
of the subgroups are larger than their velocity dispersions, then one will
typically observe redshift separation. If the galaxy associations are young
mergers, one has to expect bulk velocities which could well
exceed the above given values of velocity dispersions of subgroups; old
mergers, on the contrary, will have more overlappings.
Thus, the problem of the reconstruction of the bulk motions within the clusters
(Gurzadyan and Rauzy 1997),
particularly, the determination of the ratio $(bulk velocity)/(velocity
dispersion)$ for the galaxy associations can be important for
revealing their nature.
These issues also are topics of further studies with more data.

Thus, galaxy associations can have essential role in
the understanding
of the mechanisms of the formation of clusters of 
galaxies, since 
obviously, they should have primordial nature. By 
analogy with stellar associations and
globular clusters which have primordial stellar
population,
the problem of the study of the population of galaxies 
of the 
galaxy associations, is therefore, arising.

Observational, theoretical and numerical study of galaxy
associations,
in particular, their morphological features,  the 
process of their 
survival depending on their initial parameters
and those of the main cluster, seems of particular 
interest. 

We are thankful to D.Lynden-Bell and E.Escalera for valuable discussions
and the anonymous referee for valuable comments.
V. G. is grateful to G.Comte for hospitality
in Observatoire de Marseille.
During that visit V.G. was supported by French-Armenian PICS and Marseille
Provence M\'etropole.

\newpage

\section{Appendix}

In this Appendix we describe the main steps in introducing the concept
{\it degree of boundness} used in Section 3, and, hence, on S-tree
technique; for detailed account we refer to the original papers
(Gurzadyan, Harutyunyan and Kocharyan 1991, 1994; Bekarian and Melkonian
1997), and particularly to (Gurzadyan, Kocharyan 1994a).

Let us start with the 
simplest system of point particles interacting according to 
a given force law in  $R^3$.  
We wish to describe the degree of boundness of those particles
in various ways.

To accomplish this goal 
we  propose  the  following  natural  approach.  Let      
$x_1(t)$ and $x_2(t),\  t \in (-T,T)$ be the trajectories of the 
two particles when the interaction is taken into  account,
and let $y_1(t)$ and $y_2(t)$  be the  corresponding  trajectories 
when the interaction  is  ``switched off",
related to the former trajectories 
by sharing the same initial data at time zero
as the corresponding interacting particles.
  In other words $y_1(t)$ and $y_2(t)$ are 
the trajectories of free particles for which
$y_i(0)=x_i(0),\ \dot{y}_i(0)=\dot{x}_i(0),\ i=1,2$.

 Let us measure the degree of boundness
 of the two interacting particles over this time period
 by the function
\begin{equation}
    m=\max_{i=1,2}{\cal N}(x_i(\cdot)-y_i(\cdot)),
\end{equation}
i.e., the maximum ``deviation" of 
either of the particles from its corresponding free trajectory,
where this deviation is measured with respect to
the following local norm ${\cal N}$ 
on $C^\infty$ parametrized curves in $R^3$ over certain interval. 
\begin{equation}
    {\cal N}(x(\cdot))=\sup_{t\in (-T,T)}\{|x(t)|,|\dot{x}(t)|\}.
\end{equation}

Now consider balls of radius $r$ at each point of trajectories of the two
interacting particles $x_i$. The union of  these balls  will 
form the two spaces
$$
    {\cal C}_i(r)=\bigcup_{t\in (-T,T)} B_{x_i(t)}(r),\ i=1,2\ .
$$
$m$ is the minimal allowed radius such that
neither of  the free  particle trajectories
escapes from its corresponding space ${\cal C}_i(m)$.

Two particles are considered to be $\rho$-bound for $\rho>0$ if 
$m\leq\rho$.

This  is easily generalized  to any finite number of particles.
$N$ 
particles labeled by the  set of integers
${\cal A}=\{1,\ldots,N\}$
form  a $\rho$-bound cluster if the 
distance between the 
corresponding trajectories of the  system  of interacting particles and  free  
ones  is  less  than  the  maximal deviation of all of the particles:
\begin{equation}
    m=\max_{a \in {\cal A}}{\cal N}(x_a(\cdot)-y_a(\cdot))\leq\rho\ .
\end{equation}
Consider now a dynamical system characterized by the following equations:
$$
    \ddot{x}_a=f_a(x),\ x \in R^{dN},\ x_a \in R^d\ .
$$
For the problem of interest the smooth  functions  $f_a$
have the form
$$
    f_a(x)=\sum_{b=1}^{N} f_{ab}(x_a,x_b)\ ,
$$
where $f_{aa}(x_a,x_a)$ 
indicates the influence of an external field on the $a$-th particle.
              
We now define the boundness function
${\cal P}_Z(Y)$, where $Y\subset Z\subset {\cal A}$.
Consider the following two systems:
\begin{description}
  \item[I.] $$\ddot{z}_a=\sum_{b \in Z}f_{ab}(z_a,z_b),
              $$
            $$z_a(0)=c_a,\  \dot{z}_a(0)=v_a, \mbox{ where }a \in Z\ ,
               $$
  \item[II.] $$\ddot{y}_a=\sum_{b \in Y}f_{ab}(y_a,y_b)\ ,
               $$
            $$y_a(0)=c_a,\  \dot{y}_a(0)=v_a, \mbox{ where } a \in Y\ ,
               $$
\end{description}
where {\bf c} and {\bf v} are constant vectors, the initial position
and velocity, respectively, of the $a$-th particle. 
The first dynamical system represents
the  subsystem of particles in ${\cal Z}$ switching off the interactions
with external particles, while the second is the same for ${\cal Y}$.

Then for the given local norm ${\cal N}$ we take
\begin{equation}
    {\cal P}_Z(Y)=\max_{a \in Y}{\cal N}(z_a(\cdot)-y_a(\cdot)),
\end{equation}
where $z_a(t), y_a(t)$ are the solutions of the systems of equations
{\bf I} and {\bf II} respectively for some time interval $(-T,T)$.
In other words the boundness of ${\cal Y}$ in ${\cal Z}$ is the maximum 
deviation of the trajectories of its particles taking into account only
internal interactions compared to the situation when interactions with
particles in ${\cal Z}$ are also included.
Our goal is to split ${\cal A}$ into
$\rho$--subsystems,
i.e., to obtain the map $\Sigma$ for this choice of
boundness function.

 The description of the
algorithm of the construction of tree-diagram let us begin by corresponding
to the system of a matrix  $D_{ab} \ a,b \in {\cal A}$,
(with positive elements) 
which determines the degree of boundness of a pair $a$ and $b$.
     The concrete choice of $D$\footnote{A list of possible choices of $D$
with different informativeness is given in (Gurzadyan and Kocharyan 1994a,
Chapter 4.1). For example, the use of projected distances, which is equivalent
to spatial correlation functions, is evidently not enough informative in
order to speak seriously about subgroupings. The use of 'projected' energy
of gravitating particles (Serna and Gerbal 1996) is also not relevant,
since it does not contain information on the degree of interaction
(gradient), while arbitrary interpretation of simple numerical experiments
without theoretical background can be misleading; e.g. in (Gurzadyan and
Kocharyan 1994b) the non-equivalence of the physical N-body system with
its computer image for certain parameters is strongly proved.}
depends on the physical system and
its properties. For  our  purposes  we  should  inquire  into  the 
systems interacting via Newtonian gravity.
     For given $\rho$ it is convenient to correspond $D$ 
another matrix $\Gamma$ in a manner:
\begin{equation}
\begin{array}{cll}
\Gamma_{ab}=0 \mbox{ if } &D_{ab} <    \rho \mbox{ and } &D_{ba} < \rho,\\
\Gamma_{ab}=1 \mbox{ if } &D_{ab} \geq \rho \mbox{  or } &D_{ba} \geq \rho;
\end{array}
\end{equation}
    
$\Gamma$ can be considered as a matrix describing a graph with 
$\{1,\ldots,N\}$ 
apexes, with two connected apexes $a$ and $b$, if $\Gamma_{ab}=1$.
The $A$-set  of 
a $\Gamma$ would be  connected if
$\forall a',a'' \in A\  \exists a_i \in A,\  i=1,\ldots,k$ 
sequence of apexes so that  
$a_1=a', a_k =a''$ and $\forall i\  1\leq i\leq k-1\   \Gamma_{i,i+1}=1.$
 
     The definition of a  $\rho$--bound cluster
  given  above  can  be reformulated now as a set of  
corresponding  particles  $A$  being  a 
connected subgraph of the graph $\Gamma$ (equivalent the matrix)  so  that 
there is no other connected subgraph $B$ including $A:\  A \subset B.$
     If one defines {\cal P} as follows
\begin{equation}
    {\cal P}_XY=\max\limits_{\stackrel{y \in Y}{z \in X\backslash Y}}
                 \{D_{yz}\}\ ,
\end{equation}
the problem of the search of a $\rho$-bound cluster is reduced  to  that 
of a connected $\Gamma$-graph.

     Consider now what matrices $D$ one can choose,
 if $d=3$ and
$$
    M_a\ddot{x}_a^{i}=-\sum_{b \in {\cal A}\backslash \{a\}} f_{ab}^{i}, 
$$
$$    
f_{ab}=GM_aM_b \frac{x_a-x_b}{| x_a-x_b |^3},\quad  a \ne b,
$$    
$$
f_{aa}=0,
$$
where $M_a$ is the mass of $a$-th particle and
$$
    | x_a |=\left(\sum_{i=1}^{3}\left[x_a^i\right]^{2}\right)^{1/2}.
$$
     One can show that to account for the information on the
dynamics of the system one can choose the
matrix $D$ using the Riemannian curvature of the phase space of the
system (Arnold 1989).   
     It is well known that the Hamiltonian system by means of 
Maupertuis principle can be represented as a  geodesic  flow  in  a 
Riemannian $3\cdot N$ dimensional manifold with a defined metric. 
The  deviation of the geodesics  of  the 
configurational space and hence the properties of  the  Hamiltonian 
system are determined by Riemannian  tensor  of  curvature $Riem$
via Jacobi equation:
\begin{equation}
   \nabla_{\bf u}\nabla_{\bf u}{\bf n}+Riem({\bf n},{\bf u}){\bf u}=0,
\end{equation}
where {\bf u} is the tangential vector to the geodesic, 
{\bf n} is a deviation vector.

In some basis the Jacobi equation can be written as
$$
   \ddot n^{\mu}+K_{\nu}^{\mu}u^{\nu}=0,
$$
where $u^\mu=\langle{\bf e}^\mu,{\bf u}\rangle$, 
$n^\mu=\langle{\bf e}^\mu,{\bf n}\rangle$, 
$R^\mu_{\lambda\nu\rho}=\langle{\bf e}^\mu,
Riem({\bf e}_\nu,{\bf e}_\rho){\bf e}_\lambda\rangle$.
For a system  of $N$  gravitating  particles  the  K and 
Riemannian tensor has the form (Gurzadyan and Savvidy 1984, 1986)
$$
    K^{\mu}_{ \nu} = Riem^{\mu}_{ \lambda \nu \rho}u^{\lambda} u^{\rho}  = 
$$
$$
- \frac{1}{2W}
      [\delta^{\mu}_{\nu} W_{\lambda\rho} u^{\lambda} u^{\rho} +
 W^{\mu}_{\nu} u^2 - u^{\mu} W_{\lambda\nu} u^{\lambda} 
            - u_{\nu} W^\mu_\lambda u^{\lambda}] -
$$
$$
\frac{3}{4W^2}
      [(u^{\mu} W_{\nu} 
      - \delta^{\mu}_{\nu} W_{\lambda} u^{\lambda}) W_{\rho} u^{\rho} +  
 (u_{\nu} W_{\lambda} u^{\lambda}  
      - W_{\nu} u^2 W^{\mu}] -
$$
$$
\frac{1}{4W^2}
     [\delta^{\mu}_{\nu} u^2 - u^{\mu} u_{\nu}]\|dW\|^2, 
$$
where
$$
R^i_{ab}=x^i_a-x^i_b,\quad R_{ab}=| R^i_{ab} |,
$$
$$
W=E+\sum_{a=1}^{N}{\displaystyle\sum_{b=1}^{a-1}}G\frac{M_aM_b}{R_{ab}},\\
$$
$$
u^{\mu}=u^{(a,i)}=\frac{\dot{x}^i_{a}}{\sqrt2W},\\
$$
$$
g_{\mu\nu}=M_a\delta_{\mu\nu}\\
$$
$$
u_{\mu}=g_{\mu \nu} u^{\nu}, u^2=u^{\nu} u_{\nu} =1/W.
$$
The derivatives of $W$ have the following form
$$
W_{\mu}=W_{(a,i)}=-{\displaystyle
        \sum\limits_{\stackrel{c=1}{c\ne a}}^{N}}
        G\frac{M_aM_cR_{ac}^i}{R_{ac}^3},
$$
$$
W_{\mu \nu}=W_{(a,i)(b,j)}=G\frac{M_aM_b}{R_{ab}^3}
            \left(\delta_{ij}-\frac{3R_{ab}^iR_{ab}^j}{R_{ab}^2}\right),
            \mbox{ if } a\ne b,
$$
$$
W_{\mu \nu} =W_{(a,i)(a,j)}=-{\displaystyle
        \sum\limits_{\stackrel{c=1}{c\ne a}}^{N}}
           G\frac{M_aM_c}{R_{ac}^3}
             \left(\delta_{ij}-\frac{3R_{ac}^iR_{ac}^j}{R_{ac}^2}\right),
$$
$$
\|dW\|^2 ={\displaystyle\sum_{a=1}^{N}\sum_{i=1}^{3}}W^2_{(a,i)}/M_a,
$$
$$
W_{\mu} u^{\mu} 
={\displaystyle\sum_{a=1}^{N}\sum_{i=1}^{3}}W_{(a,i)}u^{(a,i)},\\
$$
$$
W_{\mu \nu} u^{\mu}u^{\nu}
  ={\displaystyle\sum_{a=1}^{N}\sum_{i=1}^{3}\sum_{b=1}^{N}\sum_{j=1}^{3}}
  W_{(a,i)(b,j)}u^{(a,i)}u^{(b,j)},
$$
$$
W_{\mu \nu} u^{\nu} ={\displaystyle\sum_{b=1}^{N}\sum_{j=1}^{3}}
W_{(a,i)(b,j)}u^{(b,j)}.
$$

From here we have
\begin{equation}
    D_{ab}=\max_{i,j}\left\{-K^{\mu}_{ \nu},0\right\},
\end{equation}
where  $K$  is the curvature and the  system 
consisting of the all $a$ and $b$ pairs is reduced to a geodesic flow in a 
Riemannian 3N-dimensional manifold.

The algorithm of computer analysis is described in 
(Gurzadyan et al 1994). Therefore here we briefly mention
the  procedure  of 
its operation and the ways of  presentation 
of the obtained results via S-tree diagrams.
    
 Consider a system of $N$ particles  of  masses  $M_a$ with  given 
coordinates $x_a$ and velocities $\dot{x}_a$.  By  first 
step we find out the multifunction $\Sigma$ for concrete  matrix  $D$.
     As we showed above, for given $D$ and $\rho$ the problem is  reduced 
to the search of connected parts  of  the  graph $\Gamma(\rho)$.
  In  general, however, for subsystems defined  by
${\cal P}$ another  algorithm  can  be used.

     Thus we use an algorithm which finds out the  connected  parts 
of the graph $\Gamma$ for given $\rho$, but simultaneously obtains all  those 
parts for all $\rho$. 
     In other words, given the parameters
$$
    M_a,\ x_a(0),\ \dot{x}_a(0),\  a \in {\cal A}
$$
and the matrix $D$ the algorithm finds out the function $\Sigma$.

     The latter can be represented in a form  of  tree-diagram (S-tree),  so 
that its height is determined by the value of $\rho$. For each value  of 
the latter the sets
$$
    \{{\cal A}_1(\rho),\ldots,{\cal A}_d(\rho)\},
$$
are constructed in a way  that  each  line  $L_a$ corresponds  to  a 
certain particle. As a result one has  a  tree-diagram  indicating 
the set  $\{ L_a : a \in {\cal A}\}$.

The computer experiments on toy models using the described technique are
described in (Gurzadyan et al 1991, 1994). For further development of S-tree
technique see (Bekarian, Melkonian 1997).
\vspace{0.2in}

\newpage
\noindent
{\it Figure caption}.
\vspace{0.1in}

The redshift histograms of the main clusters (solid 
line) with indicated
subgroups (dashed and doted lines) for a sample of Abell 
clusters studied
by S-tree method.

\label{reconstruction}

\end{document}